\documentclass{phb-proc4-auth}

\usepackage{graphicx}
\usepackage{amssymb}

\begin{document}
\begin{frontmatter}

\journal{SCES '04}

\title{Weak itinerant ferromagnetism in YCo$_9$Si$_4$}

\author[1]{H. Michor\corauthref{x}}
\author[1]{M. El-Hagary}
\author[1]{S. \"Ozcan}
\author[1]{A. Horyn}
\author[1]{E. Bauer}
\author[1]{M. Reissner}
\author[1]{G. Hilscher}
\author[2]{S. Khmelevskyi}
\author[2]{P. Mohn}
\author[3]{P.Rogl}
\address[1]{Institut f\"ur Festk\"orperphysik, T.U. Wien, A-1040 Wien, Austria}
\address[2]{Center for Computational Materials Science, T.U. Wien, A-1040 Wien, Austria}
\address[3]{Institut f\"{u}r Physikalische Chemie, Universit\"{a}t Wien, A-1090 Wien, Austria}
\thanks[A]{The work was supported by the Austrian Science Foundation Fonds,
P-15066-Phy. M. El-H. is on leave from Helwan University, Cairo, Egypt. 
A.H. is on leave from Ivan Franko Lviv National University, Lviv, Ukraine,
with support from \"OAD.}
%

\corauth[x]{Corresponding Author: 
Email: michor@ifp.tuwien.ac.at}

\begin{abstract}

Weak itinerant ferromagnetism in YCo$_9$Si$_4$ below about 25\,K is studied by means of
magnetisation, specific heat, and resistivity  measurements.
Single crystal X-ray Rietveld refinements at room temperature reveal a fully ordered 
distribution of Y, Co and Si atoms within the tetragonal space group {\sl I4/mcm} isostructural
with LaCo$_9$Si$_4$. The latter exhibits itinerant electron metamagnetism with an 
induced moment of about 1\,$\mu_B$/f.u.\ above 6\,T, whereas YCo$_9$Si$_4$
exhibits a spontaneous magnetisation $M_0\simeq 12$\,Am$^2$/kg at 2\,K which 
corresponds to an ordered moment of about 1.6\,$\mu_B$/f.u.\  indicating weak itinerant 
ferromagnetism. 

\end{abstract}


\end{frontmatter}

Recent interest on weak itinerant ferromagnetism  
e.g.\ in ZrZn$_2$~\cite{pfleiderer} in the context with quantum critical phenomena 
motivated the search for new materials showing weak itinerant ferromagnetism or 
being close to a ferromagnetic (FM) instability. 
An interesting system in this respect is the solid
solution LaCo$_{13-x}$Si$_x$ where ferromagnetism vanishes near
the stoichiometric composition LaCo$_9$Si$_4$~\cite{El-Hagary} 
where full translational symmetry (space group $I4/mcm$) is
confirmed by single crystal X-ray diffractometry~\cite{mi_LCS}.
LaCo$_9$Si$_4$ is a strongly exchange enhanced Pauli paramagnet and  
exhibits an itinerant electron metamagnetic phase transition at about 3.5\,T for $H||c$ 
and 6\,T for $H\bot c$, which is the lowest value ever found for rare earth 
intermetallic compounds~\cite{mi_LCS}. In this paper, we report on low
temperature measurements on the isostructural and isoelectronic compound YCo$_9$Si$_4$
which was initially reported in Refs.~\cite{skolozdra,gorelenko}  
to be FM with $T_C\simeq 848$\,K.
   
Polycrystalline samples of YCo$_9$Si$_4$ were synthesized by 
induction melting of pure elements (Y 3N, Co 4.5N, Si 6N) under protective 
argon atmosphere and subsequent annealing at 1050$^{\circ }$C for one week. 
The phase purity and composition has been verified by means of electron
microprobe analysis. The crystal structure was determined by means of 
single crystal X-ray diffraction ($R_{F^2}=$\,2\%) revealing a fully ordered distribution 
of Y, Co and Si atoms with the LaFe$_9$Si$_4$-type~\cite{tang} 
with a single rare earth site, three cobalt sites and again a single Si site.
The lattice parameters are $a=7.754(1)$\,\AA\ and  $c=11.487(1)$\,\AA\ 
at room temperature (see Ref.~\cite{mi_LCS} for experimental details). 
Crystallographic order is also corroborated by a reasonably low residual resistivity 
$\rho_0=7\mu\Omega$cm (see below).

\begin{figure}
\centering
\includegraphics[width=0.92\columnwidth]{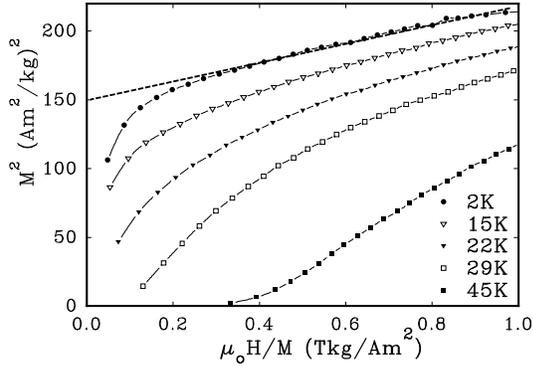}
\caption{Arrott plot $M^2$ vs. $H/M$ of isothermal magnetization data of YCo$_9$Si$_4$.}
\end{figure}

Temperature and field dependent magnetisation  measurements $M(T,H)$ on YCo$_9$Si$_4$
depicted in Fig.~1 as an Arrott plot $M^2$ versus $H/M$ reveal weak ferromagnetism below 
about 25\,K with an extrapolated spontaneous magnetisation $M_0\simeq 12$\,Am$^2$/kg at 2\,K
(see the dashed line in Fig.~1) corresponding to 1.6\,$\mu_B$/f.u. and 
a longitudinal susceptibitity in the FM regime,
$\chi_0\sim 0.25$\,Am$^2$/kgT, determined from the $\mu_0H/M$ axis intercept of the dashed line 
extrapolation in Fig.~1. 
The Curie-temperature $T_C$ is around 25\,K
in reasonable agreement with specific heat and transport anomalies (see below). 

\begin{figure}
\centering
\includegraphics[width=0.92\columnwidth]{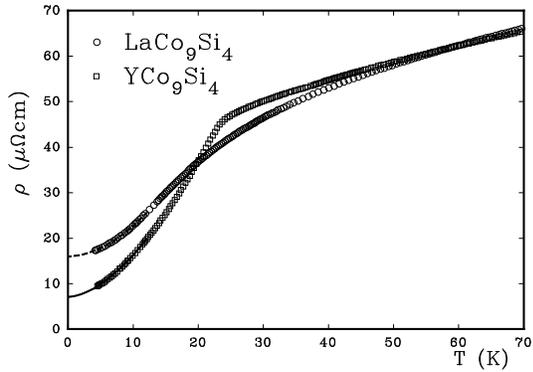}
\caption{Temperature dependent resistivity $\rho(T)$ of YCo$_9$Si$_4$ and LaCo$_9$Si$_4$;
dashed and solid lines are fits, see text.}
\end{figure}

The temperature dependent resistivity $\rho (T)$ of YCo$_9$Si$_4$ shown in Fig.~2
reveals a significant change of slope around about 25\,K which is associated with
the onset of ferromagnetism. Below about 15\,K, in the FM state,
$\rho (T)$ is well described by a power law behavior $\rho (T)=\rho_0+AT^{\alpha}$
(see the solid line in Fig.~2) with $\rho_0=7\mu\Omega$cm, $A= 0.176\mu\Omega$cm/K$^{-\alpha}$ 
and $\alpha = 1.72$. 
The corresponding fit for nearly ferromagnetic LaCo$_9$Si$_4$ (dashed line in Fig.~2) 
yields $\rho_0=16\mu\Omega$cm, $A= 0.085\mu\Omega$cm/K$^{-\alpha}$ and $\alpha = 1.9$
indicating a spin fluctuation (Fermi liquid) regime for the latter compound. 

\begin{figure}
\centering
\includegraphics[width=0.92\columnwidth]{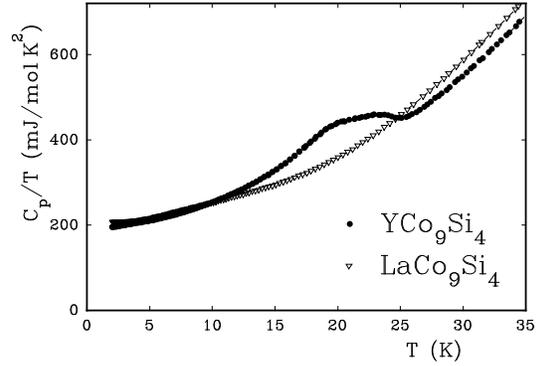}
\caption{Temperature dependent specific heat as  $C/T$ vs. $T$ of YCo$_9$Si$_4$ and LaCo$_9$Si$_4$.}
\end{figure}

The specific heat of YCo$_9$Si$_4$ and (for comparison) LaCo$_9$Si$_4$ is shown in Fig.~3 as 
$C/T$ vs. $T$ revealing for both compounds a relatively large electronic Sommerfeld value 
$\gamma$ close to 200\,mJ/mol\,K$^2$ and in the case of YCo$_9$Si$_4$ a small somewhat
broadened anomaly associated with the second order phase transition towards weak itinerant 
ferromagnetism with a jump $\Delta C/T$ of the order of 100\,mJ/mol\,K$^2$ in approximate  
agreement with the Stoner-Wohlfarth model (see e.g.\ Ref.~\cite{mohn}) yielding 
$\Delta C/T_C= M_0^2/\chi_0T_C^2\sim 70$\,mJ/mol\,K$^2$.
In the case of exchange enhanced Pauli paramagnetic LaCo$_9$Si$_4$ the value of
$\gamma\simeq 200$\,mJ/mol\,K$^2$ can be compared with the density of states obtained 
from {\sl ab-initio} electronic structure calculations, $N(E_F)\sim $19\,states/eV\,f.u., 
revealing a spin-fluctuation mass enhancement $\lambda_{spin}\sim 3.3$~\cite{mi_LCS}.

For YCo$_9$Si$_4$ band calculations have been performed in the same manner as 
described in Ref.~\cite{mi_LCS} for LaCo$_9$Si$_4$ yielding practically the same
picture with respect to the Co $d$-bands as for LaCo$_9$Si$_4$ and  
within the numerical accuracy the same density of states at the Fermi level.
The spin-fluctuation mass enhancement $\lambda_{spin}$ is thus very similar in 
YCo$_9$Si$_4$ and LaCo$_9$Si$_4$.   
Band calculations at the experimental lattice constant yield a FM ground state 
for both compounds,
which is experimentally confirmed only for YCo$_9$Si$_4$ while LaCo$_9$Si$_4$ shows a paramagnetic
ground state and metamagnetism.
In analogy to the conclusions drawn for LaCo$_9$Si$_4$ we expect also for YCo$_9$Si$_4$
in the FM state the largest moments of about 0.3--0.4\,$\mu_B$/Co to be at the 
$16k$ Co-sites and significantly smaller moments at the $4d$ and $16l$ Co-sites.

\end{document}